%% file: main.tex
\documentclass[conference]{IEEEtran}
\IEEEoverridecommandlockouts

\usepackage{cite}
\usepackage{amsmath,amssymb}
\usepackage{graphicx}
\usepackage{booktabs}
\usepackage{multirow}
\usepackage{siunitx}
\usepackage{xcolor}
\usepackage{url}
\usepackage[hidelinks]{hyperref}
\usepackage{textcomp}
\usepackage{balance}
\usepackage{acronym}
\usepackage[caption=false,font=footnotesize]{subfig}

\input{acronyms.tex}
\acrodef{isac}[ISAC]{Integrated Sensing and Communication}
\acrodef{lif}[LIF]{Leaky Integrate-and-Fire}

\graphicspath{{figs/}}

\begin{document}

\title{Keeping Inference in the Radio: Design and Measurement of a Spiking Accelerator for mmWave ISAC Activity Recognition}

\author{%
\IEEEauthorblockN{Anthony S\'anchez\IEEEauthorrefmark{1},
Rafael Ruiz\IEEEauthorrefmark{1},
Jos\'e A. Belloch\IEEEauthorrefmark{2},
Jes\'us O. Lacruz\IEEEauthorrefmark{1}\IEEEauthorrefmark{3},
Joerg Widmer\IEEEauthorrefmark{1}%
\thanks{\IEEEauthorrefmark{3}Corresponding author: jesusomar.lacruz@networks.imdea.org.}%
\thanks{This work is supported by the Horizon Europe Research and Innovation programme (SNS JU) under Grant Agreements No.~101292662 (IoT-ZERO) and No.~101192521 (MultiX), and by the Comunidad de Madrid through the DISCO6G-CM project (TEC-2024/COM-360).}}
\IEEEauthorblockA{\IEEEauthorrefmark{1}IMDEA Networks Institute, Madrid, Spain\\
\{anthony.sanchez, rafael.ruiz, jesusomar.lacruz, joerg.widmer\}@networks.imdea.org}
\IEEEauthorblockA{\IEEEauthorrefmark{2}Universidad Carlos III de Madrid, Legan\'es, Spain\\
jbelloc@ing.uc3m.es}
}
\maketitle

\begin{abstract}
Integrated sensing and communication allows wireless networks to recognize human activity from the channel estimates that the receiver computes anyway. The inference, however, runs on the general-purpose processors of the base station, access point or handset, which are the first components a device switches off to save energy. Always-on sensing therefore requires a classifier that is small enough to remain powered while the rest of the system sleeps. We investigate whether Spiking Neural Networks (SNNs) can provide such a classifier. To this end, we generate SNN accelerators for micro-Doppler activity recognition on the public DISC dataset (IEEE~802.11ay at \SI{60}{GHz}) with the open-source Spiker+ framework, and implement them together with a spike encoder on a ZCU208 RFSoC, so that the complete chain from micro-Doppler frame to activity label runs in programmable logic. A fully connected SNN with \num{132} neurons achieves \SI{94.7}{\percent} accuracy (\SI{90.7}{\percent} macro-F1) on unseen subjects and performs an inference in \SI{206}{\micro\second}. Compared to conventional networks of the same topology on the same device, it requires an eighth of the logic and \num{3.4}$\times$ less dynamic energy per decision than a network fed the same binary spikes, at comparable accuracy, and \num{7.6}$\times$ less than a network fed the 16-bit spectrogram, at \num{1.2} points lower accuracy. Quantization-aware training reduces its weights to 3~bits within half a percentage point of the 6-bit network at the same encoder threshold.
\end{abstract}

\section{Introduction}
\label{sec:intro}

\begin{figure}[t]
\centering
\includegraphics[width=0.88\columnwidth]{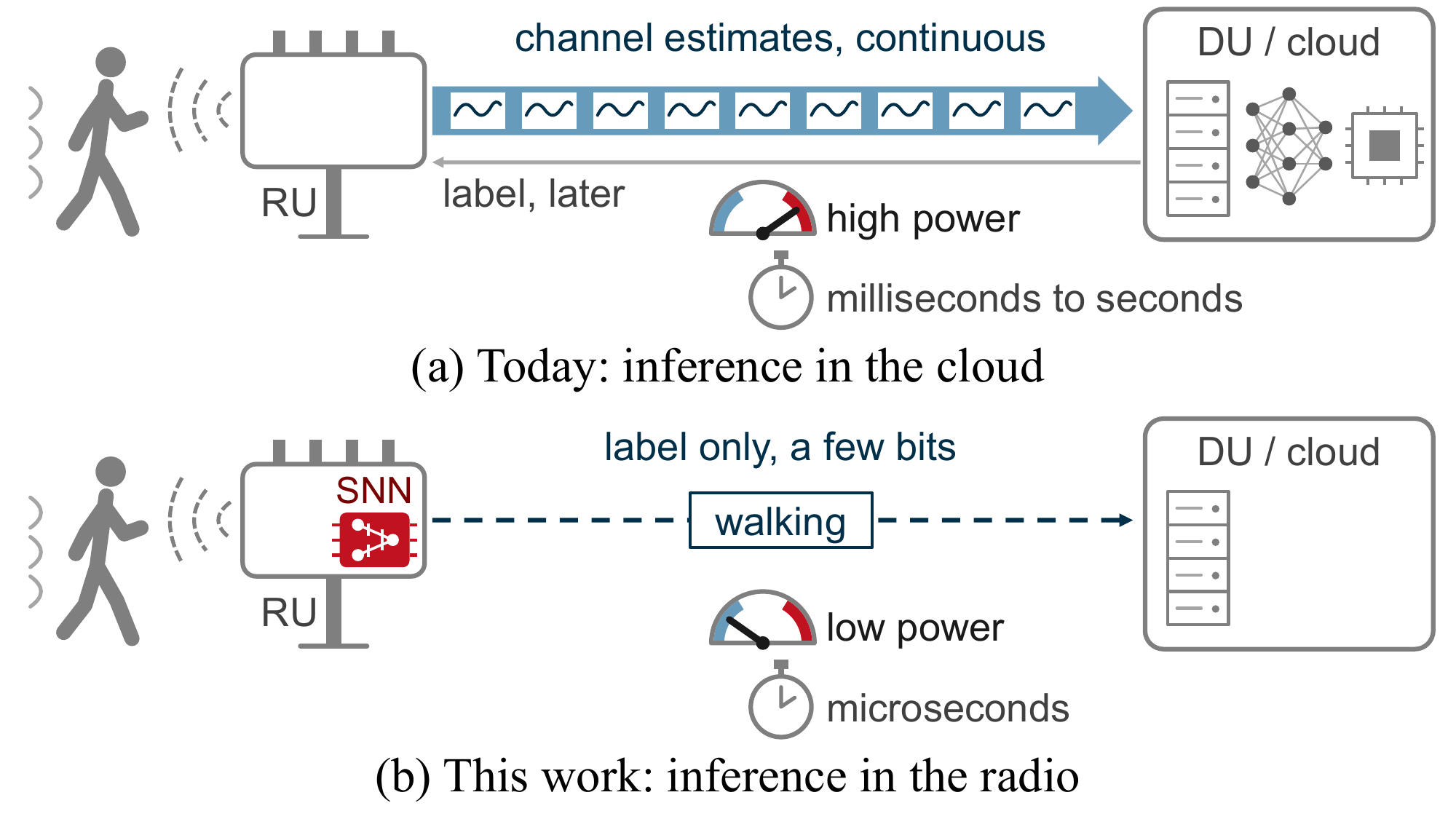}
\vspace{-2mm}
\caption{Sensing inference (a) in the cloud, fed by a continuous stream of channel estimates, and (b) in the \acf{ru}, exporting only the label.}
\label{fig:motivation}
\vspace{-4mm}
\end{figure}

\ac{isac} turns wireless links into sensors, since the channel estimates that a receiver computes for data reception change when a person moves. The inference that interprets these estimates, however, runs on the general-purpose part of the system, i.e., on the processors of the Wi-Fi \ac{ap}~\cite{meneghello2023bf} or the handset, on the \ac{du} of an \ac{oran} base station~\cite{doro2022dapps}, or in the cloud (Fig.~\ref{fig:motivation}a). While the system is active, this is the natural place for it.

Sensing, however, is most useful when it never switches off, and these processors are the first components that a device puts to sleep. An always-on sensing function therefore requires a dedicated low-power path, similar to the block that detects a wake word while a phone sleeps. Such a path consists of a reduced receiver that keeps estimating the channel, a stage that extracts the \ac{md} signature, i.e., the Doppler spectra that the moving limbs and torso imprint on successive channel estimates~\cite{pegoraro2023rapid}, and a classifier that detects a few critical events, such as a fall or a call for help, and then wakes up the rest of the system. Since the classifier exports only a few bits per decision, neither the link nor the processor at the other end has to stay awake. The main challenge is thus a classifier that is small enough to remain powered and completes each decision within the observation interval at very low power.

\acp{snn} are a promising candidate for such a classifier. They communicate through binary events, so that a synaptic operation is an accumulation rather than a multiply--accumulate~\cite{eshraghian2023training}, and their membrane state integrates the \ac{md} frames as they arrive, with a few hundred state variables instead of the buffered window of a \ac{cnn}. \acp{snn} have been studied for radar and RF sensing~\cite{safa2023snngesture,cicciarella2024learned}, where they can even outperform conventional networks~\cite{yu2026freqmatch}, but few of these results have been measured on hardware. Yu et al.~\cite{yu2026freqmatch} measure latency on a neuromorphic chip but estimate energy theoretically, \acp{snn} for activity recognition on communication-derived channel data have only been evaluated in software~\cite{cicciarella2026sparse}, and NeuroRadar~\cite{zheng2023neuroradar} designs a dedicated spiking radar front end. To the best of our knowledge, no work implements a spiking classifier in the hardware of a communication radio.

In this paper, we analyze whether a small \ac{snn} can serve as such an always-on classifier. We train a fully connected \ac{snn} with \num{132} neurons on the public DISC dataset~\cite{pegoraro2025disc}, generate fixed-point accelerators with the open-source Spiker+ framework~\cite{carpegna2024spikerplus}, and measure their accuracy, latency, resource usage and power on an AMD Xilinx \ac{rfsoc}, with a quantized conventional network of the same topology and input as reference. Our main contributions are:

\begin{enumerate}[\setlength{\IEEElabelindent}{0pt}]
\item To the best of our knowledge, the first hardware measurements of \ac{snn} inference for \ac{isac} \ac{har} on communication-derived channel data, on a transceiver-class \ac{rfsoc}.
\item A reusable integration of Spiker+ cores, with a streaming \acs{axi} wrapper that provides cycle-exact latency instrumentation and state reset between observations, and a spike encoder for the \ac{md} frames.
\item A comparison of post-training and quantization-aware training, which shows that 6-bit weights preserve the float32 accuracy and that quantization-aware training brings 3-bit weights to within half a point of the 6-bit network, with half the weight memory and \SI{23}{\percent} less logic.
\item A same-device comparison with non-spiking networks of the same topology, which shows that the \ac{snn} requires an eighth of the logic and \num{3.4}$\times$ less dynamic energy per decision than a network on the same binary input at comparable accuracy, and \num{7.6}$\times$ less than a network on the 16-bit input at \num{1.2} points lower accuracy.
\end{enumerate}

\section{From Packets to Micro-Doppler}
\label{sec:system}

\subsection{Sensing as a By-Product of Communication}
\begin{figure*}[t]
\centering
\subfloat[Processing chain]{\includegraphics[height=1.5in]{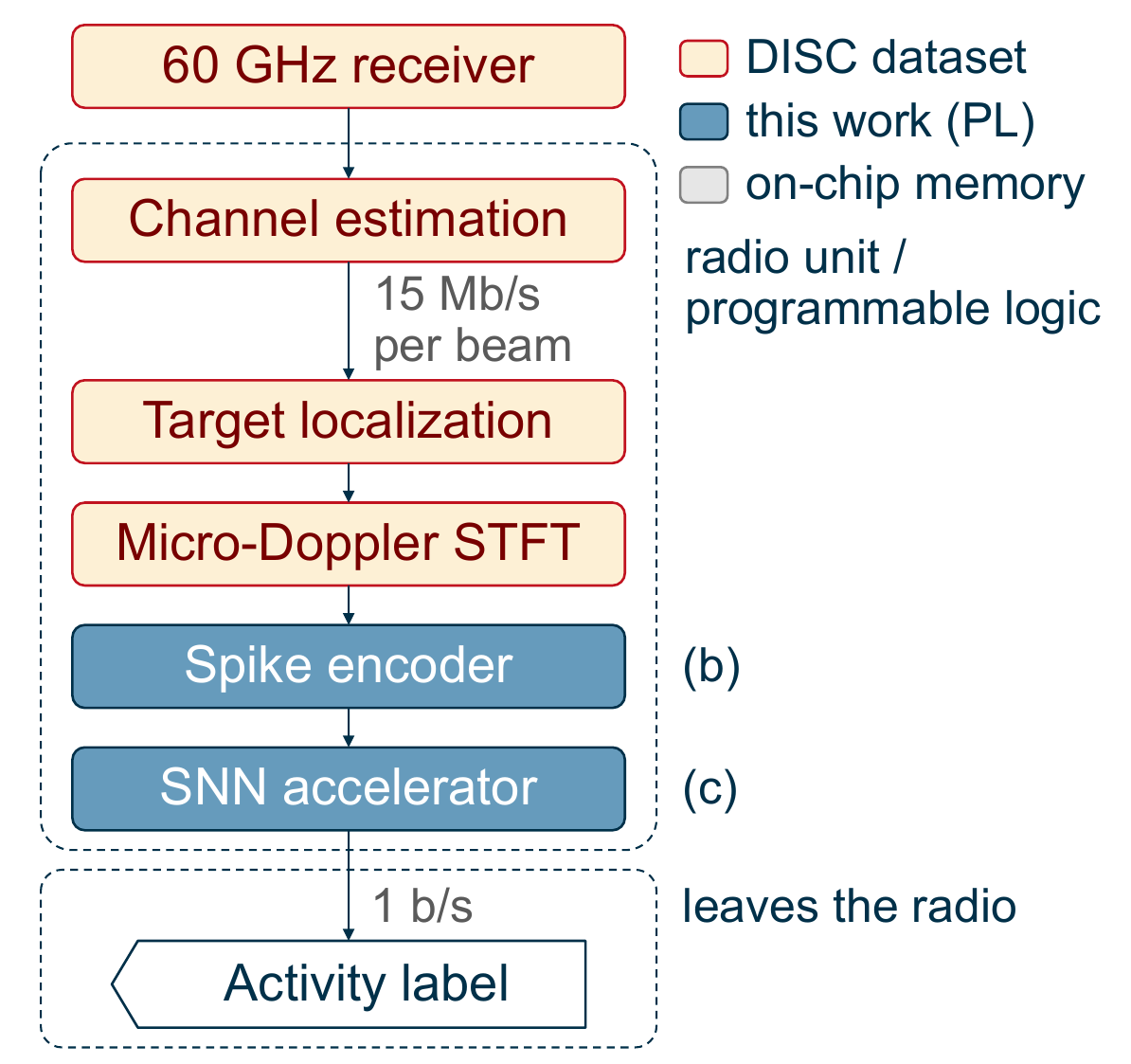}\label{fig:system-a}}\hfil
\subfloat[Spike encoder]{\includegraphics[height=1.5in]{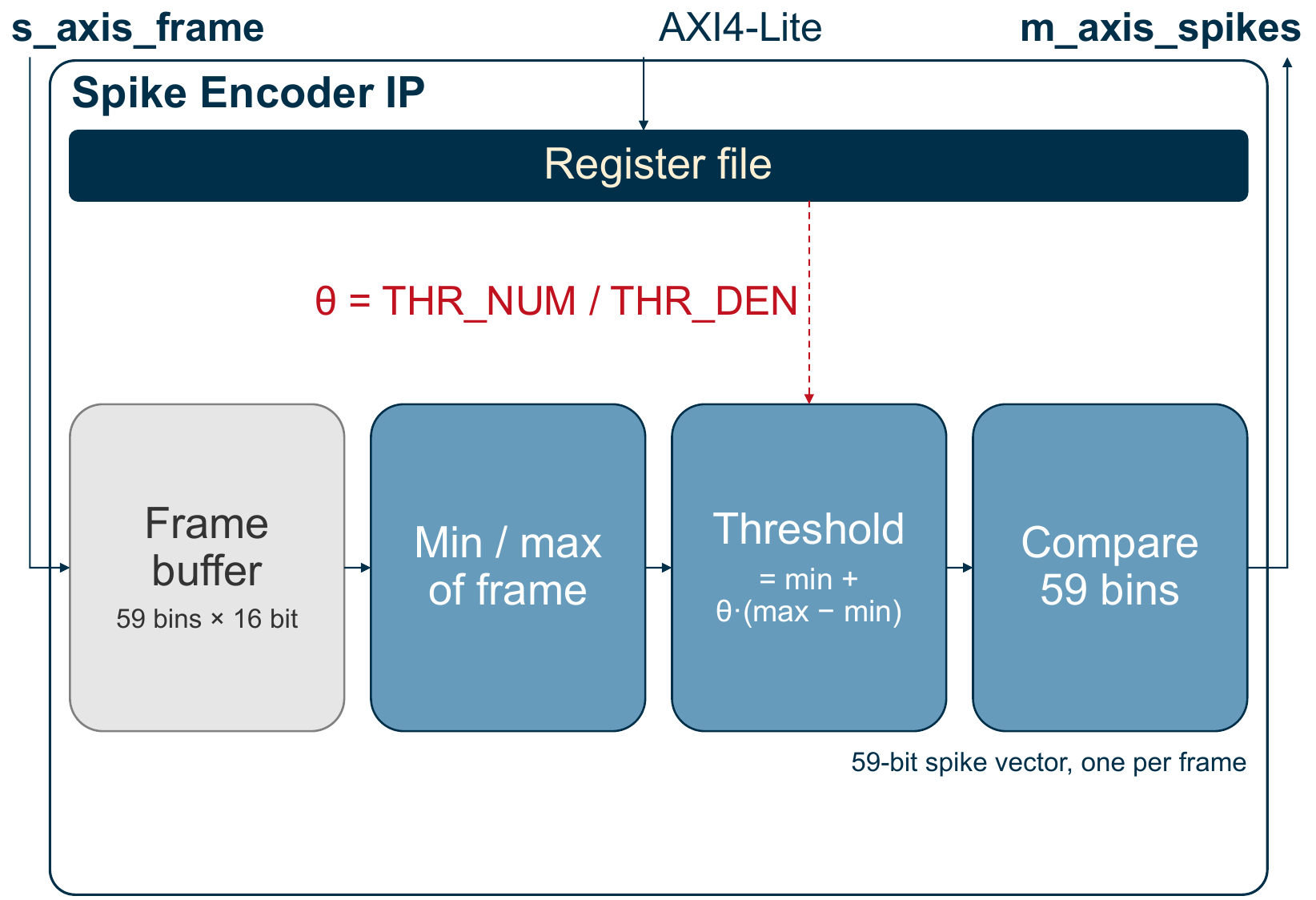}\label{fig:system-b}}\hfil
\subfloat[SNN accelerator]{\includegraphics[height=1.5in]{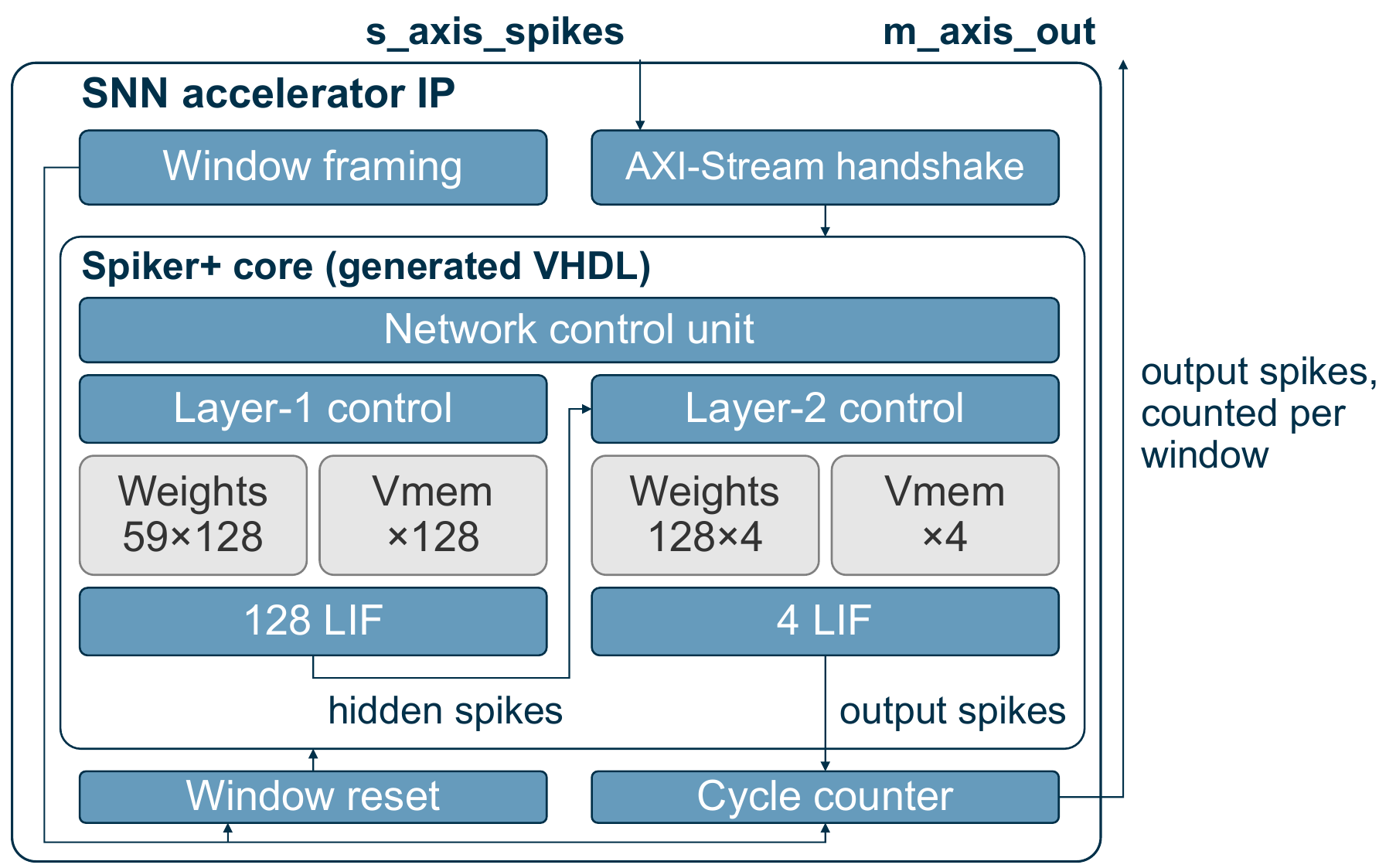}\label{fig:system-c}}
\caption{(a) Processing chain, from the 802.11ay receiver to the activity label; orange blocks come from the DISC dataset, blue and grey ones are implemented in this work. (b) Spike encoder. (c) \ac{snn} accelerator: the Spiker+ core inside the \ac{axi} stream wrapper.}
\label{fig:system}
\vspace{-2mm}
\end{figure*}

In an \ac{isac} receiver, the sensing input is a by-product of data reception. The receiver estimates the channel from the pilots or training fields of each packet, and when a person moves, the taps that correspond to reflections off the body change from one estimate to the next. A \ac{stft} over successive estimates of these taps converts this variation into the \ac{md} spectrogram (Fig.~\ref{fig:system-a}). Throughout the paper, a \emph{frame} denotes one column of this spectrogram, i.e., the Doppler spectrum of one \ac{stft} segment, and a \emph{window} denotes the sequence of consecutive frames that the classifier receives as one input. In DISC~\cite{pegoraro2025disc}, IEEE~802.11ay packets at \SI{60}{GHz} are transmitted every \SI{0.27}{ms}, which yields an unambiguous velocity range of \SI{\pm 4.5}{m/s}, and the 64-point \ac{stft} provides a resolution of \SI{0.14}{m/s}, sufficient to separate the torso of a walking person from the swing of the limbs. Consecutive \ac{stft} segments overlap by 32 packets~\cite{pegoraro2025disc}, so the radio produces one frame every \SI{8.64}{ms}.

Exporting the raw channel estimates is costly at \ac{mmwave} frequencies. In DISC-A, each packet carries one training unit, i.e., about \num{3700} channel estimates per second, which with 128 complex taps and 16~bits per real and imaginary part amount to \SI{15}{Mb/s} per beam; a link that sweeps twelve beams for localization, as in DISC-B, produces \SI{180}{Mb/s}. If the classifier runs in the radio (Fig.~\ref{fig:motivation}b), only a two-bit label per window, about \SI{1}{b/s}, leaves the device, seven orders of magnitude less than a single beam.

\subsection{Problem Formulation}
The radio produces one \ac{md} frame every \SI{8.64}{ms}, and the classifier outputs one activity label per window of $N = 200$ frames, i.e., every \SI{1.73}{s}, one to two gait cycles of a walking person and far more often than a person changes activity. The classifier must process each frame as it arrives instead of storing the window, so that its memory depends on the number of neurons rather than on the $N \times 59$ values of a window, and it must complete each decision well within the \SI{1.73}{s}. The design itself does not depend on these values, since it processes one frame at a time and the window length is a parameter.

\subsection{Dataset and Protocol}
\begin{figure}[t]
\centering
\includegraphics[width=0.98\columnwidth]{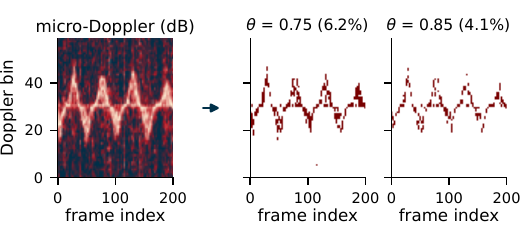}
\vspace{-3mm}
\caption{Spike encoding of a waving-hands window of 200 frames (\SI{1.73}{s}). Left: normalized \ac{md} spectrogram (dB). Right: spike rasters after the per-frame relative threshold at $\theta = 0.75$ and $\theta = 0.85$ (\SI{6.2}{\percent} and \SI{4.1}{\percent} of the bins active in this window).}
\label{fig:binarization}
\end{figure}

We use DISC-A, the part of DISC~\cite{pegoraro2025disc} recorded with a single forward-pointing beam. It contains 433 sequences, more than one hour of \ac{cir} data in which seven subjects walk, run, sit down/stand up and wave hands (Fig.~\ref{fig:binarization} shows the latter), together with the \ac{md} spectrograms distributed with the dataset. The evaluation is subject-disjoint: subjects 1--5 train the network and subjects 6 and 7 test it, which measures the generalization to people never seen in training. In contrast to the DISC benchmark, which validates on subject 6 and tests on subject 7, we use no validation subject, so the fixed-point configurations and the encoder threshold reported below were chosen on the test subjects. Windows in which a walking or running subject is momentarily still, detected as a lack of motion energy away from the static return, are discarded in both sets (\SI{2}{\percent} of the test windows). The training set is balanced to \num{1341} windows per class by replicating minority-class windows with additive Gaussian noise ($\sigma = 0.1$), whereas the \num{812} test windows (200 frames, cut with a step of 50) keep their natural distribution (\SI{68}{\percent} walking).

\section{Accelerator Design and Measurement Method}
\label{sec:design}
We generate the accelerators with Spiker+~\cite{carpegna2024spikerplus}, an open-source framework that takes an \ac{snn} described in snnTorch, trains it, quantizes weights and membrane potentials to fixed point and generates synthesizable VHDL, so that the hardware remains a function of the trained model. We chose a \ac{fpga} over a neuromorphic chip, since the latter would add a second device with its own package, power domain and interfaces to the radio. The \ac{md} frames are produced where the channel is estimated: in the baseband of an \ac{ap} or a handset, or, in a cellular network with an \ac{oran} 7.2x split, in the \ac{du}, which is precisely the unit that sleeps, so the always-on path of Fig.~\ref{fig:motivation}b needs a reduced channel estimator in the \ac{ru} ahead of the classifier. In either case the classifier consumes the frames without any data leaving the chip, and its VHDL can be ported to the baseband ASIC.

\subsection{Spike Encoding}
\label{sec:encoding}
A spiking network {operates on} binary events, so the spectrogram has to be {converted} into spikes{,} either by an encoder or by feeding the values as input {currents into the first layer, which gives up the multiplier-free arithmetic~\cite{eshraghian2023training}. Rate and latency coding require} several time steps per value{,} and delta coding {discards the steady components. Since} the \ac{md} spectrogram already has a time axis, we {instead map each} frame {to one time step and let} a Doppler bin spike if its magnitude is large compared {to} the rest of the frame.

The encoder (Fig.~\ref{fig:system-b}) determines the minimum and maximum of the bins of a frame and emits a spike for every bin above $\min + \theta\,(\max - \min)$. Because the reference is the frame's own range, the rule is independent of the absolute signal level, e.g., of gain and distance, which is what allows the encoder to sit directly behind the \ac{stft} of a radio; it requires no memory of earlier frames, but the frame maximum always fires, so a frame without motion also produces spikes. The DISC spectrograms are distributed in dB and already normalized per frame to $[0,1]$~\cite{pegoraro2025disc}, so on this data the rule coincides with a fixed threshold on the unsigned 16-bit values the encoder receives. We use $\theta = 0.75$ for SNN-S, SNN-M and SNN-L, which keeps the torso and limb traces (Fig.~\ref{fig:binarization}) with \SI{6.4}{\percent} of the bins active over the test set.

\subsection{Network and Training}
\label{sec:training}
The \ac{snn} has \num{59} inputs, one per Doppler bin, since the distributed spectrograms keep 59 of the 64 \ac{stft} bins, a hidden layer of \num{128} first-order \ac{lif} neurons and \num{4} output \ac{lif} neurons, one per activity, and runs one time step per \ac{md} frame, so that it operates on a stream without buffering a window. All neurons fire at a threshold of 1.0 and decay with $\beta = 1-2^{-4} = 0.9375$, so that the leak is a 4-bit arithmetic shift. Hidden neurons use subtractive reset, output neurons do not reset within a window, and all potentials are cleared between windows. In float32, $\beta$ gives the membrane a memory of about 16 frames; in fixed point, however, the leak $\lfloor v/16 \rfloor$ vanishes below $2^{4-f}$, i.e., below the threshold for $f = 4$ (SNN-S and SNN-SS), so the hidden neurons of those two configurations do not leak at all, and only SNN-L leaks approximately as the float32 model does. The label is the class whose output neuron fires most over the window. All networks are trained offline with backpropagation through time and surrogate gradients~\cite{eshraghian2023training} (arctangent with $\alpha = 2$, Adam, learning rate $5\times10^{-4}$, batch size 64, \num{15} epochs) with a cross-entropy loss on the output spike counts. SNN-S, SNN-M and SNN-L share one float32 training run at $\theta = 0.75$ and differ only in their post-training quantization; this run is the best of three seeds (mean accuracy \SI{93.1}{\percent} $\pm$ 1.7), selected on the test subjects, which makes the float32 accuracy and the hardware accuracies derived from it optimistic by about 1.5 points. SNN-M at $\theta = 0.85$ is a separate training run at that threshold, and SNN-SS is trained with the quantization in the loop (Section~\ref{sec:quant}).

\begin{table}[t]
\caption{SNN fixed-point configurations.}
\label{tab:config}
\centering
\footnotesize
\begin{tabular}{@{}lcccc@{}}
\toprule
Fixed-point configuration & SNN-S & SNN-M & SNN-L & SNN-SS \\
\midrule
Weight bits      & 6 & 6 & 8 & 3 \\
Membrane bits    & 7 & 8 & 10 & 6 \\
Fractional bits (both) & 4 & 6 & 8 & 4 \\
Quantization     & post & post & post & aware \\
Encoder $\theta$ & 0.75 & 0.75 & 0.75 & 0.85 \\
\bottomrule
\vspace{-6mm}
\end{tabular}
\end{table}

\subsection{Fixed-Point Quantization and HDL Generation}
\label{sec:quant}
Spiker+ quantizes weights and membrane potentials to two's-complement fixed point with $f$ fractional bits. We swept weight widths of 6--10 bits, membrane widths of 7--10 bits and $f$ of 4--10 bits (\num{140} combinations) and selected, among the most accurate in Spiker+'s emulation on the test subjects, three configurations of different width from one set of float32 weights (Table~\ref{tab:config}), for which Spiker+ generated the VHDL. The deployed designs include a one-line correction of the output-neuron reset in the VHDL that Spiker+~2.0.3 generates, which we have reported upstream
.

The accuracy that Spiker+ reports for its software emulation of the quantized network differs from that of the generated hardware by up to several points, because the emulation does not apply saturation, leak and spike propagation in the order of the circuit. We therefore built a bit-exact software model of the generated core, verified against the board, and use it to explore configurations without a bitstream for each; all hardware accuracies in this paper are nevertheless measured on the board, with the on-chip encoder.

Each layer updates all its neurons in parallel while feeding the input spikes serially (Fig.~\ref{fig:system-c}), so the first layer spends 59 cycles per time step and the second 128, and a window costs roughly $200\times(59+128)$ cycles plus control overhead. A layer skips any time step without input spikes, but with our encoder the first layer never does, since the frame maximum always fires; the latency is therefore almost independent of the input.

To reduce the bit-width further, we also trained with the quantization in the loop, using a forward pass with a differentiable copy of the accelerator's fixed-point arithmetic, so that the network learns under the same rounding, saturation and spike timing as the hardware. This quantization-aware training is not part of Spiker+, but it produces ordinary weights that deploy through the same flow.
We use it for SNN-SS, with 3-bit weights and the encoder threshold $\theta = 0.85$ (Table~\ref{tab:config}).

\subsection{System Integration on the RFSoC}
The Spiker+ cores expose a bespoke start/ready handshake and raw spike vectors rather than a standard interface. We therefore wrapped the core in an \ac{axi} stream interface, so that any \ac{axi} master can push frames at any rate, with back-pressure in both directions. The wrapper also frames the stream into windows, clears all membrane potentials at the end of each window, and stamps the cycle count onto the last beat, which yields a cycle-exact latency independent of the host. Four counters and a comparator in the wrapper count the output spikes of a window and return the label.

The spike encoder is a streaming block in front of the wrapped \ac{snn} (Fig.~\ref{fig:system-b}). It receives one 59-bin frame of 16-bit values over its own \ac{axi} stream, emits the spike vector within \num{161} clock cycles and takes $\theta$ over \ac{axi}-Lite. Implemented together with the wrapper and the \ac{snn} on the board, it reproduces the software encoder bit for bit on all \num{812} test windows, and the labels match the bit-exact model. PYNQ\footnote{\url{https://www.pynq.io}} manages the data streaming and the registers, and an \ac{axi} \ac{dma} moves the frames from the processor to the \ac{pl} and the labels back. In a radio, the \ac{stft} of the digital front end, followed by a logarithm since the encoder expects dB values, would replace the \ac{dma}.

\subsection{Measurement Method}
\textbf{Latency} is measured in two ways. Inside the \ac{pl}, as in a radio, the wrapper's cycle counter gives the processing time of a window. Through the processor, as in the prototype, a timer around the \ac{dma} transfer gives the end-to-end time seen by software (median over \num{1000} passes). \textbf{Resources} are the post-implementation utilization reports of Vivado 2023.1 on the XCZU48DR, for the \ac{snn} cores implemented out of context and for the ANN and encoder instances of the complete design; for SNN-M the two differ by \SI{3}{\percent}.
\textbf{Power} is that of the VCCINT rail, the supply of the \ac{pl}, computed as the product of the current and bus voltage read from the board's INA226 monitor through the Linux hwmon interface (\SI{1}{mA} resolution). Because one current reading takes longer than one inference, the test window is stored on chip and replayed through the encoder to the accelerator back-to-back, so that the core runs continuously while the \num{1000} readings per point are taken, each averaging over tens of inferences.

\subsection{Non-Spiking Reference}
\label{sec:baseline}
As reference, we implemented the same topology without spikes, \texttt{Linear(59,128)}, ReLU, \texttt{Linear(128,4)}, applied to each frame with the 200 logit vectors summed before the decision, in two versions: ANN, fed with the same binarized spectrogram as the \ac{snn}, and ANN16, fed with the 16-bit normalized spectrogram. Both are trained in float32 with the recipe of Section~\ref{sec:training}, quantized to int8 weights and 16-bit activations, generated with hls4ml\footnote{\url{https://fastmachinelearning.org/hls4ml/}} (Vitis HLS 2023.1) with 192 multipliers, which Vivado maps to 200 and 192 DSP slices, wrapped with the same stream contract as the \ac{snn} and measured with the same method. Neither closes timing at \SI{250}{MHz}, so all designs are compared at \SI{187.5}{MHz}. While topology and input are controlled, hls4ml generates a parallel datapath with DSP multipliers and Spiker+ a serial one with memories, so the differences in logic reported below include the implementation style; in particular, the binary ANN uses the same multipliers as ANN16, since hls4ml does not exploit 1-bit inputs.

\section{Results}
\label{sec:results}

\subsection{Classification Accuracy}
Table~\ref{tab:acc} gives the recall of every class, the accuracy and the macro-F1, which weighs the four activities equally, whereas accuracy is dominated by the \SI{68}{\percent} of walking windows. SNN-S, SNN-M and SNN-L derive from the same float32 model (\SI{94.6}{\percent} accuracy, \SI{91.8}{\percent} macro-F1). SNN-L matches it, SNN-M keeps the accuracy and loses one point of macro-F1, and SNN-S loses three points of accuracy and nine of macro-F1. We attribute the latter to its four fractional bits, with which the hidden neurons do not leak below threshold (Section~\ref{sec:training}): SNN-SS has the same leak but was trained under it, whereas SNN-S was quantized after a float32 training with the full leak. In all post-training configurations, walking gains recall while running and waving lose it. Raising $\theta$ from \num{0.75} to \num{0.85} removes \SI{35}{\percent} of the input spikes and \SI{25}{\percent} of the hidden spikes, and costs the retrained SNN-M \num{1.5} points of accuracy. At that threshold, quantization-aware training makes 3-bit weights usable: SNN-SS reaches \SI{92.7}{\percent}, half a point below SNN-M, whereas the same network quantized after training reaches \SI{87.8}{\percent} (five seeds, bit-exact model).

The non-spiking networks (Section~\ref{sec:baseline}) measure the cost of binarization and of spiking separately. Fed the 16-bit spectrogram, ANN16 reaches \SI{95.9}{\percent} and \SI{92.8}{\percent} macro-F1; fed the same spikes as the \ac{snn}, the ANN drops to \SI{93.6}{\percent} and \SI{88.2}{\percent}, so binarization costs this architecture \num{2.3} points of accuracy and \num{4.5} of macro-F1. SNN-M, on the same spikes, sits in between: \num{1.1} points above the binary ANN and \num{1.2} below ANN16, differences of 9 and 10 windows that lie within the spread between training seeds ($\pm 1.7$ points), so we consider the three accuracies comparable. The \ac{cnn} of the DISC benchmark reaches a mean per-class F1 of \SI{96.1}{\percent}~\cite{pegoraro2025disc} and an \ac{snn} with a learned encoder on the raw \acp{cir} a macro-F1 of \SI{95.8}{\percent}~\cite{cicciarella2026sparse}, both on subject 7 alone; on that subject SNN-L reaches \SI{92.1}{\percent} accuracy and \SI{91.1}{\percent} macro-F1, five points behind networks with convolutions and 3.5 to 30 times more parameters.

\begin{table}[t]
\caption{Per-class recall, accuracy and macro-F1 on the test set.}
\label{tab:acc}
\centering
\footnotesize
\setlength{\tabcolsep}{3pt}
\begin{tabular}{@{}clcccccc@{}}
\toprule
 & & \multicolumn{4}{c}{Recall (\%)} & Acc. & Macro- \\
\cmidrule(lr){3-6}
 & & Walk & Run & Sit/St. & Wave & (\%) & F1 (\%) \\
\midrule
\multirow{3}{*}{SW}
 & FC SNN   & 95.5 & 89.0 & 98.9 & 89.7 & 94.6 & 91.8 \\
 & FC ANN   & 97.6 & 85.7 & 85.7 & 83.3 & 93.6 & 88.2 \\
 & FC ANN16 & 97.5 & 93.4 & 85.7 & 98.7 & 95.8 & 92.7 \\
\midrule
\multirow{7}{*}{HW}
 & SNN-S                  & 99.1 & 68.1 & 96.7 & 57.7 & 91.4 & 82.8 \\
 & SNN-M                  & 98.9 & 78.0 & 96.7 & 82.1 & 94.7 & 90.7 \\
 & SNN-L                  & 98.6 & 81.3 & 97.8 & 84.6 & 95.2 & 91.8 \\
 & SNN-M, $\theta$ 0.85   & 96.6 & 79.1 & 95.6 & 83.3 & 93.2 & 88.7 \\
 & SNN-SS, $\theta$ 0.85  & 93.3 & 94.5 & 95.6 & 83.3 & 92.7 & 88.1 \\
 & FC ANN, int8           & 98.4 & 82.4 & 86.8 & 80.8 & 93.6 & 88.2 \\
 & FC ANN16, int8         & 99.1 & 91.2 & 93.4 & 82.1 & 95.9 & 92.8 \\
\bottomrule
\end{tabular}
\vspace{-3mm}
\end{table}

\subsection{Latency}
With the encoder in front of it, the accelerator processes a stored window in \num{38676} clock cycles in all \ac{snn} configurations, i.e., \SI{206}{\micro\second} at \SI{187.5}{MHz}, the clock at which all designs are compared. The count does not vary between repetitions and changes by less than \SI{1}{\percent} between windows, since only the second layer skips steps. The ANN needs \num{16802} cycles (\SI{89.6}{\micro\second}) and ANN16 \num{19616} (\SI{104.6}{\micro\second}). Through the processor, the \ac{dma} and the software add \SI{225}{\micro\second} per window, which disappear when the frames arrive from the \ac{pl}. On a live stream, the encoder and the accelerator spend about 354 cycles (\SI{2}{\micro\second}) on each \SI{8.64}{ms} frame as it arrives, so the label is available about \SI{2}{\micro\second} after the last frame and the hardware computes during about \SI{0.02}{\percent} of the time.

\begin{figure*}[t]
\centering
\begin{minipage}[t]{0.335\textwidth}
\vspace{0pt}\centering
\makeatletter\def\@captype{table}\makeatother
\caption{Post-implementation utilization of the accelerators alone (XCZU48DR).}
\label{tab:res}
\footnotesize\renewcommand{\arraystretch}{1.3}
\setlength{\tabcolsep}{2.5pt}
\begin{tabular*}{\linewidth}{@{\extracolsep{\fill}}lrrrrrr@{}}
\toprule
 & \multicolumn{4}{c}{SNN} & \multicolumn{2}{c}{ANN} \\
\cmidrule(lr){2-5}\cmidrule(l){6-7}
 & S & M & L & SS & bin & 16 \\
\midrule
LUTs      & 4095 & 4529 & 5184 & 3504 & 34374 & 30131 \\
Registers & 1824 & 1940 & 2204 & 1688 & 28891 & 27306 \\
BRAM      & 6.5  & 11   & 14.5 & 5.5  & 14    & 14 \\
DSPs      & 0    & 0    & 0    & 0    & 200   & 192 \\
\bottomrule
\end{tabular*}
\end{minipage}\hfill
\begin{minipage}[t]{0.32\textwidth}
\vspace{0pt}\centering
\includegraphics[height=1.4in]{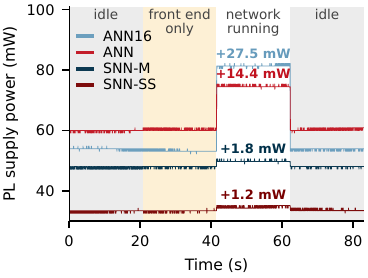}
\vspace{-3mm}
\caption{{\ac{pl} supply power across the four measurement phases (medium window, \SI{187.5}{MHz}). Labels: mean increase over idle.}}
\label{fig:powtrace}
\end{minipage}\hfill
\begin{minipage}[t]{0.30\textwidth}
\vspace{0pt}\centering
\includegraphics[height=1.4in]{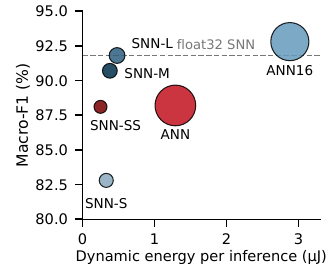}
\vspace{-3mm}
\caption{Macro-F1 on the board against dynamic energy per inference at \SI{187.5}{MHz}; bubble area proportional to LUTs.}
\label{fig:space}
\end{minipage}
\vspace{-2mm}
\end{figure*}

\subsection{Resources}
Table~\ref{tab:res} lists the utilization of the accelerators alone, without \ac{dma} and processor interconnect. The \ac{snn} cores occupy \SIrange{3.5}{5.2}{k} LUTs, \SIrange{1.7}{2.2}{k} registers, \numrange{5.5}{14.5} BRAM tiles and no DSP slice, since multiplying a weight by a spike reduces to adding the weight or not and the leak is a shift, and every resource grows with the bit-width. The ANN and ANN16 need \numrange{6.7}{7.6}$\times$ the LUTs, \numrange{14}{15}$\times$ the registers and \numrange{192}{200} DSPs of SNN-M, at a comparable number of BRAM tiles; the 16-bit input costs no logic, since hls4ml implements both with the same multipliers. The 3-bit SNN-SS needs \SI{23}{\percent} fewer LUTs and half the BRAM of SNN-M. The encoder adds \num{611} LUTs and \num{965} registers, and SNN-M with its encoder uses \SI{1.2}{\percent} of the LUTs of the ZCU208, a footprint that fits on far smaller devices.

\subsection{Power and Energy}
Fig.~\ref{fig:powtrace} shows the power of the \ac{pl} supply through four phases, i.e., idle, front end alone (encoder and replay memory), network running, and idle again, for four of the designs. With the bitstream loaded and nothing running, the rail draws \SIrange{33}{50}{mW} with the \ac{snn} designs and \SIrange{54}{60}{mW} with the ANNs. Part of this difference is a real cost of the ANNs, whose 14 to 15 times more registers load the clock tree also when idle; the rest follows the placement, since the clock buffers of every region that holds logic draw power whether it toggles or not, and the spread among the \ac{snn} designs tracks the clock regions each bitstream occupies (7 to 17) and its BRAM tiles. We therefore compare only the increase over idle within one measurement, where the idle level is stable to \SI{0.4}{mW}. The encoder adds nothing the monitor can resolve. ANN16 adds \SI{27.5}{mW} and the ANN \SI{14.4}{mW}: with binary inputs, most multiplier inputs are zero and the datapath toggles half as much. SNN-S adds \SI{1.6}{mW}, SNN-M \SI{1.8}{mW}, SNN-L \SI{2.3}{mW} and SNN-SS \SI{1.2}{mW}. Per inference, SNN-M thus needs \SI{0.38}{\micro\joule}, \num{3.4}$\times$ less than the ANN (\SI{1.29}{\micro\joule}) and \num{7.6}$\times$ less than ANN16 (\SI{2.88}{\micro\joule}), and SNN-SS \SI{0.25}{\micro\joule}. At one inference every \SI{1.73}{s}, the dynamic energy of SNN-M amounts to \SI{0.2}{\micro\watt} of average power, so in an always-on path the cost of the classifier is set by its static power, i.e., by its area, and by the receiver and \ac{stft} ahead of it, which this paper does not quantify.

\subsection{Design Space}
Fig.~\ref{fig:space} places the six accelerators in the macro-F1--energy--area plane. ANN16 is the most accurate and the most expensive point, at \num{7.6}$\times$ the energy and \num{7}$\times$ the LUTs of SNN-M. SNN-M and SNN-L sit above the binary ANN in macro-F1 and SNN-SS matches it, at a seventh to a tenth of its LUTs and \numrange{2.7}{5.2}$\times$ less energy. Among the \ac{snn} configurations, area follows the bit-width, and quantization-aware training moves SNN-SS to the lower-left corner of the plane.

\section{Discussion}
\label{sec:discussion}
\textbf{What limits accuracy.} On subject 7, SNN-L (\SI{91.1}{\percent} macro-F1) remains below the \SI{96}{\percent} that the DISC \ac{cnn}~\cite{pegoraro2025disc} and an \ac{snn} with a learned encoder and 28k parameters~\cite{cicciarella2026sparse} reach. ANN16 shows where the gap lies: on the same topology, binarizing the input costs \num{4.5} points of macro-F1, of which SNN-M recovers half on the same spikes, possibly through its temporal integration, although the difference lies within the spread between seeds; the rest separates a fully connected network with 8k parameters and no temporal feature extraction from the convolutional designs.

\textbf{Where the savings come from.} Since the encoder always fires the frame maximum, the latency is almost constant and the dynamic power changes by less than \SI{0.5}{mW} with the input activity. The savings over the ANNs thus come from accumulate-only arithmetic, serial processing and small area rather than from input sparsity, consistent with hardware-aware studies that find \acp{snn} more energy-efficient than dense networks only at low spike rates~\cite{dampfhoffer2023}. An absolute noise floor in the encoder would make frames without motion, the most common input of an always-on sensor, also the cheapest ones.

\textbf{Energy and area.} At one decision every \SI{1.73}{s}, \SI{0.38}{\micro\joule} corresponds to \SI{0.2}{\micro\watt}, far below the idle power of any \ac{fpga} design and of the front end that the always-on path keeps running, i.e., receiver, ADC, channel estimation and \ac{stft}, which we have not quantified. What matters in a product is the standby power of the block, which in an ASIC is dominated by leakage and scales with area. There, the \ac{snn} needs no multipliers, while its weight memory equals that of a dense network of the same width, and quantization-aware training halves it at a cost of half a point.

\textbf{Scope.} DISC-A covers four activities by seven subjects on one monostatic link. Always-on applications need even fewer classes, since presence, a fall or a call for help are one or two critical classes against background activity, and each additional class costs one output neuron without changing the latency.

\section{Conclusion}
We have shown that the inference of a micro-Doppler activity classifier can be kept in the radio. A spiking network with \num{132} neurons and its encoder run in the programmable logic of an \ac{rfsoc}, achieve about \SI{95}{\percent} accuracy on unseen subjects and perform an inference in \SI{206}{\micro\second}. Compared to conventional networks of the same topology, the \ac{snn} needs an eighth of the logic and \num{3.4}$\times$ less dynamic energy per decision than a network on the same binary input, at comparable accuracy, and \num{7.6}$\times$ less than a network on the 16-bit input, at \num{1.2} points lower accuracy; quantization-aware training reduces its weights to 3~bits at a cost of half a point at the same encoder threshold. On the prototype, the static power of the \ac{fpga} fabric dominates. In an ASIC, such a block would reside in a low-power domain whose leakage scales with area, which is where the small footprint of the \ac{snn} pays off and allows sensing to remain active while the rest of the system sleeps.

\balance
\bibliographystyle{IEEEtran}
\bibliography{refs}

\end{document}

%% file: acronyms.tex
\acrodef{prop}[\textit{SPIDR}]{SPIke-Domain Opportunistic Receiver}

\acrodef{abft}[A-BFT]{Association Beamforming Training}
\acrodef{ack}[ACK]{Acknowledge}
\acrodef{adc}[ADC]{Analog-to-Digital Converter}
\acrodef{aoa}[AoA]{Angle of Arrival}
\acrodef{aod}[AoD]{Angle of Departure}
\acrodef{ap}[AP]{Access Point}
\acrodef{amc}[AMC]{Advanced Mezzanine Card}
\acrodef{awv}[AWV]{Antenna Wave Vector}
\acrodef{axi}[AXI]{Advanced eXtensible Interface}
\acrodef{adc}[ADC]{Analog-to-Digital Converter}
\acrodef{ber}[BER]{Bit Error Rate}
\acrodef{bft}[BFT]{Beamforming Training}
\acrodef{bp}[BP]{Beam Pattern}
\acrodef{brp}[BRP]{Beam Refinement Phase}
\acrodef{bss}[BSS]{Blind Source Separation}
\acrodef{cacc}[CACC]{Cross-Antenna Cross-Correlation}
\acrodef{casr}[CASR]{Cross-Antenna Signal Ratio}
\acrodef{cs}[CS]{Compressed Sensing}
\acrodef{cdf}[CDF]{Cumulative Distribution Function}
\acrodef{cef}[CEF]{Channel Estimation Field}
\acrodef{cfar}[CFAR]{Constant False Alarm Rate}
\acrodef{cfo}[CFO]{Carrier Frequency Offset}
\acrodef{cir}[CIR]{Channel Impulse Response}
\acrodef{cfr}[CFR]{Channel Frequency Response}
\acrodef{crb}[CRB]{Cramér-Rao Bound}
\acrodef{crlb}[CRLB]{Cramér-Rao Lower Bound}
\acrodef{csi}[CSI]{Channel State Information}
\acrodef{csirs}[CSI-RS]{CSI-Reference Signal}
\acrodef{cs}[CS]{Compressed Sensing}
\acrodef{cv}[CV]{Constant Velocity}
\acrodef{cnn}[CNN]{Convolutional Neural Network}
\acrodef{cots}[COTS]{Commercial-Off-The-Shelf}
\acrodef{dac}[DAC]{Digital-to-Analog Converter}
\acrodef{dft}[DFT]{Discrete Fourier Transform}
\acrodef{dl}[DL]{Deep Learning}
\acrodef{dma}[DMA]{Direct Memory Access}
\acrodef{dmg}[DMG]{Directional Multi Gigabit}
\acrodef{dmrs}[DMRS]{Demodulation Reference Signal}
\acrodef{srs}[SRS]{Sounding Reference Signal}
\acrodef{dti}[DTI]{Data Transfer Interval}
\acrodef{eadc}[EADC]{Event-based ADC}
\acrodef{edmg}[EDMG]{Enhanced Directional Multi Gigabit}
\acrodef{ekf}[EKF]{Extended Kalman Filter}
\acrodef{elu}[ELU]{Exponential-Linear Unit}
\acrodef{fft}[FFT]{Fast Fourier Transform}
\acrodef{fmcw}[FMCW]{Frequency-Modulated Continuous-Wave}
\acrodef{fim}[FIM]{Fisher Information Matrix}
\acrodef{fov}[FOV]{Field-of-View}
\acrodef{ft}[FT]{Fourier Transform}
\acrodef{fr2}[FR2]{Frequency Range 2}
\acrodef{fpga}[FPGA]{Field Programmable Gate Array}
\acrodef{gpio}[GPIO]{General Purpose Input/Output}
\acrodef{gsps}[GSPS]{Giga-Samples per Second}
\acrodef{har}[HAR]{Human Activity Recognition}
\acrodef{ht}[HT]{High Throughput}
\acrodef{ica}[ICA]{Independent Component Analysis}
\acrodef{idft}[IDFT]{Inverse Discrete Fourier Transform}
\acrodef{if}[IF]{Intermediate Frequency}
\acrodef{ifft}[IFFT]{Inverse Fast Fourier Transform}
\acrodef{ifs}[IFS]{Inter-Frame Spacing}
\acrodef{iht}[IHT]{Iterative Hard Thresholding}
\acrodef{ista}[ISTA]{Iterative Shrinkage-Thresholding Algorithm}
\acrodef{isac}[ISAC]{Integrated Sensing And Communication}
\acrodef{iot}[IoT]{Internet of Things}
\acrodef{imu}[IMU]{Inertial Measurement Unit}
\acrodef{ir}[IR]{Impulse Radio}
\acrodef{jcs}[JCS]{Joint Communication and Sensing}
\acrodef{jm}[JM]{JUMP-MUSIC}
\acrodef{jpdaf}[JPDAF]{Joint Probabilistic Data Association Filter}
\acrodef{lo}[LO]{Local Oscillator}
\acrodef{los}[LoS]{Line-of-Sight}
\acrodef{lbm}[LBM]{Loop-Back Memory}
\acrodef{lora}[LoRa]{Long-Range wide area}
\acrodef{lambda}[LAMBDA]{Least-Squares AMBiguity Decorrelation Adjustment}
\acrodef{lif}[LIF]{Leaky Integrate and Fire Neuron}
\acrodef{ls}[LS]{Least-Squares}
\acrodef{mae}[MAE]{Mean Absolute Error}
\acrodef{mcs}[MCS]{Modulation and Coding Scheme}
\acrodef{md}[$\mu$D]{micro-Doppler}
\acrodef{mimo}[MIMO]{Multiple Input Multiple Output}
\acrodef{mip}[MIP]{Mixed-Integer Programming}
\acrodef{miqp}[MIQP]{Mixed-Integer Quadratic Program}
\acrodef{mmwave}[mmWave]{Millimeter-Wave}
\acrodef{msps}[MSPS]{Mega-Samples per Second}
\acrodef{mu}[MU]{Multiple User}
\acrodef{music}[MUSIC]{MUltiple SIgnal Classification}
\acrodef{mmse}[MMSE]{Minimum Mean-Squared Error}
\acrodef{nac}[NAC]{Normalized Auto Correlation}
\acrodef{nco}[NCO]{Numerical Controlled Oscillator}
\acrodef{nlos}[NLoS]{Non-Line-of-Sight}
\acrodef{nn}[NN]{Neural Network}
\acrodef{nls}[NLS]{Nonlinear Least-Squares}
\acrodef{ofdm}[OFDM]{Orthogonal Frequency Division Multiplexing}
\acrodef{omp}[OMP]{Orthogonal Matching Pursuit}
\acrodef{per}[PER]{Packet Error Rate}
\acrodef{pdf}[p.d.f.]{Probability Density Function}
\acrodef{phy}[PHY]{Physical Layer}
\acrodef{pl}[PL]{Programmable Logic}
\acrodef{plm}[PLM]{Physical Layer Model}
\acrodef{pov}[POV]{Point-of-View}
\acrodef{ps}[PS]{Processing System}
\acrodef{pss}[PSS]{Primary Synchronization Signal}
\acrodef{po}[PO]{Phase Offset}
\acrodef{prf}[PRF]{Pulse Repetition Frequency}
\acrodef{pbch}[PBCH]{Physical Broadcast Channel}
\acrodef{qam}[QAM]{Quadrature Amplitude Modulation}
\acrodef{ransac}[RANSAC]{Random Sample Consensus}
\acrodef{rd}[RD]{Range-Doppler}
\acrodef{rf}[RF]{Radio Frequency}
\acrodef{rfsoc}[RFSoC]{Radio Frequency System on a Chip}
\acrodef{rcs}[RCS]{Radar Cross-Section}
\acrodef{rmse}[RMSE]{Root Mean Square Error}
\acrodef{rss}[RSS]{Received Signal Strength}
\acrodef{rom}[ROM]{Read Only Memories}
\acrodef{rx}[RX]{receiver}
\acrodef{sc}[SC]{Single Carrier}
\acrodef{sdr}[SDR]{Software Defined Radio}
\acrodef{sinr}[SINR]{Signal-to-Interference-plus-Noise Ratio}
\acrodef{siso}[SISO]{Single Input Single Output}
\acrodef{sls}[SLS]{Sector Level Sweep}
\acrodef{snn}[SNN]{Spiking Neural Network}
\acrodef{snr}[SNR]{Signal-to-Noise Ratio}
\acrodef{soc}[SoC]{System on a Chip}
\acrodef{spb}[SPB]{Signal Processing Blocks}
\acrodef{sqnr}[SQNR]{Signal-to-Quantization-Noise Ratio}
\acrodef{srrc}[SRRC]{Square-Root-Raised-Cosine}
\acrodef{ssb}[SSB]{Synchronization Signal Block}
\acrodef{sss}[SSS]{Secondary Synchronization Signal}
\acrodef{ssr}[SSR]{Super Sample Rate}
\acrodef{sta}[STA]{Station}
\acrodef{std}[STD]{Standard Deviation}
\acrodef{stf}[STF]{Short Training Field}
\acrodef{stft}[STFT]{Short Time Fourier Transform}
\acrodef{su}[SU]{Single User}
\acrodef{sn}[SN]{Synaptic Neuron}
\acrodef{tf}[TF]{Time-Frequency}
\acrodef{to}[TO]{Timing Offset}
\acrodef{toa}[ToA]{Time of Arrival}
\acrodef{tdoa}[TDoA]{Time Difference of Arrival}
\acrodef{tx}[TX]{transmitter}
\acrodef{ue}[UE]{User Equipment}
\acrodef{ula}[ULA]{Uniform Linear Array}
\acrodef{usrp}[USRP]{Universal Software Radio Peripheral}
\acrodef{vht}[VHT]{Very High Throughput}
\acrodef{wlan}[WLAN]{Wireless Local Area Network}
\acrodef{wnalp}[WNALP]{Weighted Normalized Auto-correlation Linear Predictor}
\acrodef{zf}[ZF]{Zero Forcing}

\acrodef{oran}[O-RAN]{Open Radio Access Network}
\acrodef{ran}[RAN]{Radio Access Network}
\acrodef{ru}[RU]{Radio Unit}
\acrodef{du}[DU]{Distributed Unit}
\acrodef{cu}[CU]{Central Unit}
\acrodef{ric}[RIC]{RAN Intelligent Controller}